# In-situ characterization of contamination within an Atomic Force Microscope tip-sample system

J. Sánchez, L. Almonte and J. Colchero

Centro de Investigación en Óptica y Nanofísica (CIOyN), Departamento Física
Facultad de Química (Campus Espinardo),
Universidad de Murcia, E-30100 Murcia (Spain).
Corresponding Author: colchero@um.es

Atomic force microscopy is based on tip–sample interaction, which is determined by the properties of tip and sample. Unfortunately, in particular in ambient conditions the tip as well as the sample are contaminated, and it is not clear how this contamination may affect data in Atomic Force Microscopy (AFM) applications. In the present work we propose to use on the one hand AFM imaging of the cantilever chips and on the other hand multidimensional AFM spectroscopy techniques to characterize the state of contamination of the tip–sample system. We find that typically AFM cantilevers may be severely contaminated when taken from typical packaging boxes that have been opened for a long time. In addition, by acquisition spectroscopy data as a function of tip-sample voltage and tip-sample distance, we are able to determine the Hamaker constant of the system, which depends strongly on the contamination within the tip–sample system. This method allows for in–situ characterization of the tip–sample system using only AFM techniques.

**Introduction**

Atomic Force Microscopy (AFM) [1,2] is a widely used technique in different fields of science, due to its extreme resolution and its high versatility. AFM allows to obtain images of surfaces, of small systems and of adsorbed atoms and molecules on a nanometric, atomic and even sub-atomic scale in environments as different as air, liquids and vacuum. AFM operation is based on the interaction between a sharp tip (the probe) and the sample to be analyzed. To obtain images, this interaction is maintained constant by changing the normal position (z-direction) as the tip is scanned over the surface(lateral, x,y-directions); while in spectroscopy operation, the lateral position is kept constant as the normal position of the sample is varied in order to access material properties ("chemical information", thus the name spectroscopy)[3]. AFM allows not only the measurement of surface topography, but also the determination of other physical characteristics; in particular electrical[4–6] and magnetic properties[7,8].

For reliable data acquisition a well-defined and stable tip is essential. It is well known that image resolution and more generally data quality depends critically on the tip geometry as well as on its material properties. Correspondingly, many companies have devoted numerous efforts to fabricate tips that are well characterized and at the same time as sharp as possible[9]. On the other hand, even the best tip as fabricated originally may be of limited use if it does not maintain its specified properties during AFM operation[10,11].

It is generally accepted that tip degradation -either tip wear or tip contamination from the sample- is mainly induced by AFM operation [10–12]. Tip degradation depends critically on the sample to be analyzed, the medium used for AFM operation, as well as the operation mode. However, a number of works have also emphasized the importance of tip contamination even prior to tip usage. In this context, several groups have investigated the different kind of contamination that may affect the tip: organic contamination of the medium, metallic pollutants of the manufacturing processes, as well as contamination induced by the Gel-Pak containers used for chip transport and storage[13,14]. As a consequence, a variety of cleaning procedures have been proposed to recover the ideal tip as manufactured initially; ranging from less aggressive methods such as washing with sodium lauryl sulfate (SDS), gratings[15,16], alcohols like ethanol or acetone[17], ultraviolet (UV)[18] and ozone treatment[19], heating to evaporate contaminants[20,21], argon bombardment and even ultrasound; to much more aggressive methods such as piranha solution cleaning[22] and the RCA process[23,24], both being very successful for removal of organic and metal contamination. We note however that these latter cleaning methods (and maybe even some of the first ones) may modify tip radius and/or its physical and chemical properties.

Finally, almost as important as cleaning is a good characterization of tip properties, in particular the contamination state of the tip. In this context, several methods for tip characterization have been proposed, such as X-ray analysis, Raman spectroscopy, contact angle measurements, scanning and transmission electron microscopy, etc.[25,26,24,27] We note that usually all these methods require removal of the tip from the AFM setup to some other equipment, which implies on the one hand a significant time effort, and on the other hand the problem related to possible tip change during the transfer process. The ex-situ analysis of the tip may therefore be of limited use to extrapolate how the tip had been during prior AFM data acquisition. Moreover, since the tip may change quite frequently during AFM operation, ex-situ methods may be quite inadequate and unproductive for many AFM experiments. Some methods have been proposed to infer tip properties - in particular the radius- using AFM techniques, which in our opinion is the optimal approach[28–31].

As discussed above, AFM is fundamentally based on tip-sample interaction. In this context, we may interpret that the tip is one half of the system, the sample being the other half. Unfortunately, the tip is that half of the system which is not directly seen, making tip characterization using AFM techniques a non-trivial task. To formalize this idea, we recall that within the Derjaguin approximation, the tip-sample force $F_n$ is[32]

$$F_n = 2\pi R\, w(d) \tag{1}$$

where $w(d)$ is the interaction per unit area of two infinite surfaces separated by a distance $d$ and $R$ is the tip radius (more precisely: $R$ is the effective radius of the tip-sample system; $1/R=1/R_{tip}+1/R_{surf}$ with $R_{tip}$ tip radius and $1/R_{surf}$ local surface curvature). Relation (1) shows that the tip-sample force is proportional to the tip geometry (described by the tip radius) on the one hand, and by the material properties on the other (described by the surface energy *w(d)*). In the context of the present work we note that relation (1) may be understood in the following way: it separates tip-sample interaction into a term describing geometry (radius *R*) and the term *w(d)* describing the chemistry of the system.

In the present work, we propose to characterize the tip following two basic ideas. First, we will (indirectly) image the tip by assuming that, to a first approximation, the apex of the tip should have a similar degree of contamination as the rest of the cantilever and of the chip onto which the tip and cantilever are attached. Second, we will assume that by precisely measuring the tip-sample interaction as a function of tip-sample distance we can infer properties related to the surface energy as well as the contact potential, and thus to the chemistry of the tip-sample system. We will show that tip contamination severely reduces tip-sample interaction, essentially due to a lower effective Hamaker constant of the tip-sample system. This leads to less interaction signal and thus to a worse signal to noise ratio.

**Experimental**

***Data was acquired*** using Dynamic Atomic Force Microscopy (DAFM) on a Nanotec Electronica AFM system with a phase-locked loop board (PLL, bandwidth ~2 kHz), which maintained the cantilever at resonance. Images and spectroscopy were acquired using the frequency as signal for the feedback channel (Frequency Modulation dynamic mode; FM-DAFM[33]) at small oscillation, which generally implies non-contact operation (so-called attractive regime, see also[34] for more detail). Tip-sample distance is estimated to be between 5 and 10 nm, ensuring low noise imaging with high spatial resolution, not only of topography but also of electrostatic interaction (~ 20 nm). Platinum coated silicon tips ($\nu_0 \approx 70$ kHz) with a nominal force constant of 3 N/m were used. The nominal radius value for tip apex of these probes is specified as 15 nm by the manufacturer.

***Electrostatic measurements*** were performed by detecting the frequency shift (and thus the force gradient) induced by an alternating bias between tip and sample (also termed FM-detection of electrostatic force). This frequency shift is:

$$\frac{\delta \nu_{el}(x,y)}{\nu_0} = \frac{C''(d)}{2\, c_{lever}}(U_{bias} - U_{CP}(x,y))^2 \qquad (2)$$

where *C''(d)* is the second derivative of the capacitance, $\nu_0$ the (free) resonance frequency and $c_{lever}$ the spring constant of the cantilever. For a bias voltage $U_{bias}=U_{DC}+U_{AC}\sin(\nu_{elec}\, t)$ three frequency components of the electrostatic interaction are obtained from equation (1); a DC signal, a signal $U_\nu$ varying with the same frequency as the electrical modulation frequency $\nu_{elec}$, and a signal $U_{2\nu}$ varying with twice that frequency[5,6,35,36]. These signals $U_\nu$ and $U_{2\nu}$ are analyzed using lock-in techniques to obtain the electrostatic images ESFM$_\nu$ and ESFM$_{2\nu}$; the first is related to the contact potential difference $U_{CP}$ between tip and sample and the second to the capacitance of the tip-sample system. To measure contact potential images the Kelvin technique is used. In this technique the signal ESFM$_\nu$ is nullified with an auxiliary feedback system by adjusting the tip voltage, then the voltage applied to the tip is precisely the contact potential ($V_{DC} = V_{CP}$). Frequency detection gives higher spatial resolution as compared to force detection ESFM, in addition, it allows for a correct determination of contact potential (see, for example,[34,37]). An external lock-in board was employed for the ESFM and KPM measurements using $U_{AC}$ voltages as low as $U_{AC} \approx 500$ mV at an electrical modulation frequency $\nu_{elec} \approx 7$ kHz. Further details of the set-up and ESFM and KPM working operation modes are described elsewhere[34]. WSxM software was used for image processing[38]. Typically a plane filter was applied to topography images; no filter is applied to the electrostatic images.

***Multidimensional Spectroscopy: "Interaction Images".*** Multidimensional spectroscopy data is acquired as "*interaction images*" using the 3D-Mode routine of the WSXM acquisition program[38]. In this kind of spectroscopy data the horizontal axis (quick scan direction) corresponds to a voltage sweep, while the vertical axis (slow scan direction) corresponds to tip−sample distance[39]. The color scale shows the variation of resonance frequency due to tip−sample interaction as a function of bias voltage and tip-sample distance. Note that, as discussed in more detail elsewhere, for each distance, a parabola is obtained that shows a quadratic dependence of the frequency shift with bias, as expected from eq. 2 when the AFM system is operated in the noncontact regime. Fitting the experimental data to a parabola, for each (voltage) scan line three parameters are determined: the position of the minimum of each parabola, corresponding to the contact potential $U_{CP}$, the vertical "offset" of the parabola (measured interaction at the minimum of the parabola) corresponding to the Van der Waals interaction, and finally the curvature of each parabola, which is determined by the capacity, and thus by the strength of electrostatic interaction. The methodology of "interaction imaging" is described in detail in[39].

***Sample Preparation.*** For the experiments Platinum coated (on tip-side) Silicon Cantilevers (Olympus OMCL-AC240TM), silicon nitride tip-sharpened (Olympus OMCL-HA100) and All-in-One Platinum coated (on tip-side) Silicon probes (BudgetSensors AIOAl-TL) were utilized. All the experiments have been performed at room temperature and ambient conditions. For cleaning of the cantilevers the RCA procedure has been used[23]. In our experiments the two steps of the process are implemented as follows: The first step, Standard Clean-1 is performed with a solution composed of 5:1:1 parts by volume of Milli-Q Water, $NH_4OH$ (ammonium hydroxide, 29%) and $H_2O_2$ (hydrogen peroxide, 30%). When utilized as sample, the chips with the cantilevers are sonicated for about a minute in this solution, which removes organic residues. Before the next step the samples are rinsed with Milli-Q Water. The second step (Standard Clean -2) is performed with a solution composed of 5:1:1 parts by volume of Milli-Q Water, $H_2O_2$ (hydrogen peroxide, 30%), and HCl (hydrochloric acid, 37%). Again, the cantilevers are sonicated for about a minute in this solution and then rinsed with plenty of Milli-Q Water. This second step removes metallic (ionic) contaminants that may have been deposited in the Standard Clean-1 cleaning step. In addition, for silicon surfaces this second step forms a thin passivating layer. Finally, the surface of the samples is dried by blowing with $N_2$ for about 1 minute.

**Results and discussion**

**Topographic imaging of tip and flat part of a cantilever**

Figure 1 shows images where the lower side of the cantilever, that is, the side with the sensing tip, has been used as the sample. We have found that tip–sample interaction between the lateral sides of the tip cone may be huge if two plane lateral sides of the tip interact together. In a sense such a tip-probe versus tip-sample system has infinite effective tip radius R giving a huge adhesion force (recall $F_{adh}=4\pi R \gamma \cos(\varphi)$, with $\gamma$ surface energy of $H_2O$ and $\varphi$ contact angle of water on the material of the tip-sample system, usually the system "wets", then $\cos(\varphi)\approx1$ ). In this context we note that during our experiments we have observed (on an optical microscope) how the sample-tip adheres to the sensing-tip and the whole macroscopic sample (tip and chip) is moved as the probe-tip jumps to contact with the sample-tip and the whole system is moved by the scanning motion of the piezoelectric element. In spite of these problems, we have been able to image an AFM cantilever and its tip (Figure 1), which shows the flat cantilever part (Figure 1 left), the tip (Figure 1, right) and an optical image of the whole cantilever (Figure 1, middle). In the optical image the scan areas corresponding to the two AFM images are marked with two rectangles. The sample cantilever was taken out of the box and imaged without further processing (in particular no cleaning). As can be clearly observed in figure 1 the cantilever surface is covered by round islands of typically 50 nm diameter and 10-20 nm height (contact angle of these drop-like particles assuming a spherical surface: about 45°). From these topographic images we deduce that the surface of the cantilever is severely contaminated, since the images do not show the characteristic structure of evaporated, multi-crystalline platinum grains. Moreover, as observed in figure 1, essentially the same kind of contamination is observed on the flat part of the cantilever and on the sides of the tip cone. Even though we have not been able to image the tip apex at high resolution we believe that our observation supports the hypothesis that if the composition of the whole cantilever is uniform then the contamination observed on the flat part of the cantilever is analogous to the contamination present on its tip. We may assume that the contamination layer behaves like a "carpet" covering all parts of the cantilever, and in particular the tip apex used as probe in AFM applications. Therefore, if the material of the whole cantilever is uniform the contamination state of the tip can inferred by characterizing the flat part of the cantilever, which is the experimentally much simpler system.

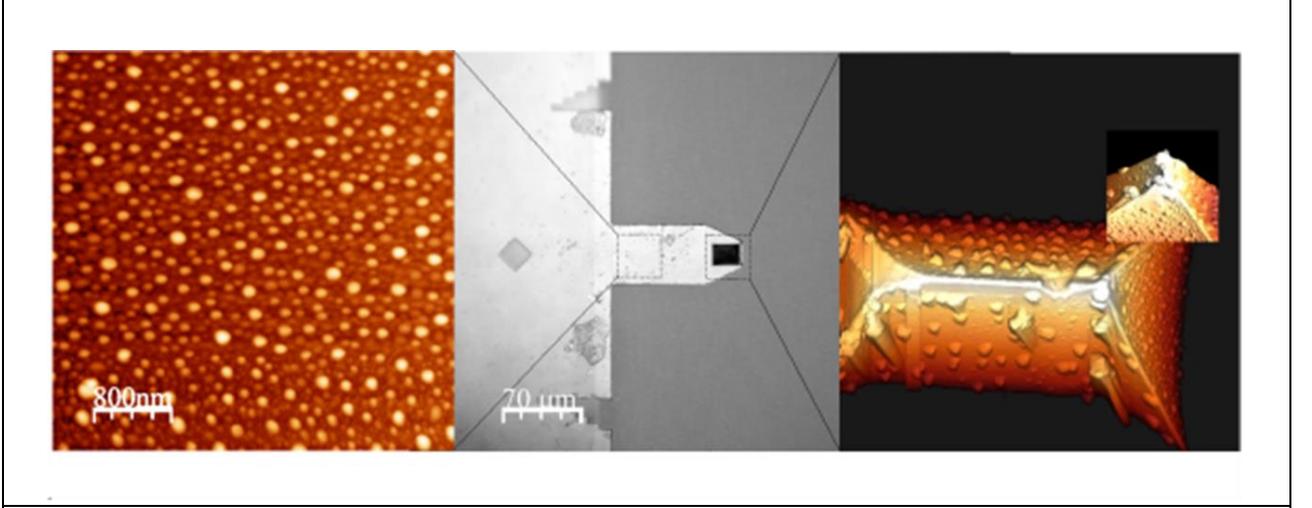

*Figure 1*

*AFM images (left and right) and optical image (middle) of the tip side of a Olympus OMCL-HA-100 AFM cantilever. Image sizes: 36x36 µm² lateral size and 25 nm grayscale for the left AFM image showing the flat region of the cantilever (left); 300x300 µm² µm for the optical image (middle); and 36x36 µm² for the right AFM image showing the flat regions of the cantilever (right). In this latter scale, the height scale shown corresponds to about 4 µm. The dotted squares in the optical image mark the two regions where the AFM images have been acquired.*

**Topography and electrostatic images of cleaned and uncleaned cantilevers: accessing material properties**

Topographic images as shown in figure 1 are quite valuable if the sample is well controlled. However, in many cases the morphology gives only a limited amount of information with respect to the chemical composition of the sample. Fortunately, AFM provides additional "secondary channels" from which compositional information can be extracted. As discussed in more detail in[34], data can be acquired in the (true) non–contact regime (nc-DAFM), where only Van der Waals and electrostatic interaction is present. In this work we will assume that the tip-sample system can be described by a metallic tip interacting electrostatically with a metallic sample and that on each metallic surface a thin dielectric film may be adsorbed (thin $\Leftrightarrow R > h/\varepsilon$). In addition, we will assume that the (second derivative of) tip-sample capacitance (see equation 2) can be approximated by the expression $C''(d) = \pi \varepsilon_0 R / (d+h/\varepsilon)^2$, where $R$ is the effective tip radius, $d$ the tip-sample distance, $h$ the (total) thickness of the dielectric films and $\varepsilon$ their (relative) dielectric constant[40–43]. For a purely metallic system were the only dielectric is air or vacuum, $h/\varepsilon \to 0$ and the normal pole $1/d^2$ of the electrostatic interaction is recovered. The total frequency shift induced by tip-sample interaction is then

$$\frac{\delta v_{el}(x,y)}{v_0} = \frac{1}{2\, c_{lever}} \left( \frac{A\,R}{3\,d^3} + \frac{\pi\,\varepsilon_0\,R}{(d+h/\varepsilon)^2} (U_{bias} - U_{CP}(x,y))^2 \right) \qquad (3)$$

where the first term with the Hamaker constant $A$ describes the Van der Waals interaction and the second term describes the electrostatic interaction. We note that the chemical composition of the sample will determine three different parameters in this relation: the Hamaker constant $A$, the pole $\cong 1/(d + h/\varepsilon)^2$ of the capacity term and the contact potential $U_{CP}$. All three terms will be used in the present work to infer the composition of the tip-sample system. In particular the Contact Potential between two materials should be quite sensitive to small variations of parameters such as the work function, and is widely used to detect adsorption layers, oxide layers, dopant concentration in semiconductors, and even temperature changes, on the sample.

Figure 2 shows images of three different experiments where the flat part of the cantilever chip has been analyzed in (true) nc-DAFM using KPM, that is, by recording the tip–sample voltage that has to be applied to have $ESFM_v = 0$. In addition to the topography, also the error signal of the feedback (frequency shift) as well as the contact potential and the signal $ESFM_{2v}$ are measured simultaneously. The three samples analyzed correspond to the surface of platinum – films evaporated onto silicon cantilevers, but with three different state of contamination: one was taken from a box that had been open for first time quite long ago (more than a year) which will be termed "*uncleaned–old*" (figure 2A, top), another one taken from a freshly open box term "*uncleaned–new*" (figure 2B, middle) and the last one, termed "*cleaned*" (figure 2C, bottom), had been cleaned just before imaging using the RCA procedure discussed in the experimental section.

We will first discuss the results obtained on the sample "*uncleaned–old*". As for the case of the cantilever imaged in figure 1 the topographic image (Fig2A(a)) shows round islands of approximately 50 nm diameter and 10-20 nm height, which we again associate to surface contamination. We note that although the surface composition of the cantilever imaged in figure 1 (material: silicon nitride) and that imaged in figure 2A (material: Pt-coated silicon) is very different, the topography of the surfaces looks quite similar. This seems reasonable taking into account that both cantilevers came from a box that had been open long time before the experiments. Therefore these cantilevers had enough time to be exposed to ambient air (each time a cantilever is extracted from the box) and become severely contaminated.

From the topography image one would conclude that the platinum surface is covered by some contamination (round islands); that is, one would expect two kind of materials on the surface: the platinum substrate and some other material –most probably organic– on top of the platinum. Surprisingly, the contact potential image is completely homogeneous, which leads to the conclusion that either the islands are completely transparent to the measurement of contact potential (more precisely: no charge transfer due to differences in work function between the platinum and the material of the island) or that the platinum surface is covered everywhere by the same material, that is, the contact potential image detects the same material on the islands as well as on the flatter part between the islands. Since the contact potential image is very sensitive to differences in work function and since it seems unlikely that platinum has the same work function as the (unknown) material of the island, we think that the most logical explanation for the homogeneous contact potential image is the second one: the whole surface is covered by the same material which corresponds to an adsorbed contamination layer. Moreover, this explanation is also supported by results to be discussed below.

As compared to the contact potential, the capacity image ($EAFM_{2\nu}$) shows a clear contrast between the lower region and the islands. According to equation 3 the capacity is $C''(d) \sim (d+h/\varepsilon)^{-2}$, we therefore interpret that the lower capacity on the islands is due to a larger thickness of the (dielectric) contamination film. Note that if the islands were composed of a conducting material, then $h = 0$ and we would expect no contrast of the capacity $C''(d)$. More precisely: feedback is performed on the Van der Waals interaction $\sim d^{-3}$; which would also keep constant the capacity $\sim (d+h/\varepsilon)^{-2}$ if $h = 0$.

Figure 2B shows the results obtained for the cantilever of type *uncleaned–new* which, as discussed above, corresponds to a cantilever which had been taken from a freshly opened cantilever box. In this case, all images – topography, contact potential and capacity – look quite homogeneous. Unfortunately, these homogeneity does not indicate whether the surface is homogeneously clean, or homogeneously contaminated.

Finally, figure 2C shows images corresponding to a cantilever cleaned using the RCA method. The topography image is, on a larger scale (200nm) quite flat showing a small irregular corrugation (5 nm) which is typical of the silicon substrate on which the thin platinum – film (nominally 20 nm thickness) is evaporated. On a smaller scale (image not shown) this sample shows the typically rounded structure of the poly crystalline platinum grains ($\approx$ 20 nm size). A few higher structures (3 nm) can be recognized in the topographic images. Interestingly only one of these higher structures gives a contrast in the contact potential image (-90 mV as compared to 70 mV for the substrate), otherwise the contact potential image is completely homogeneous.

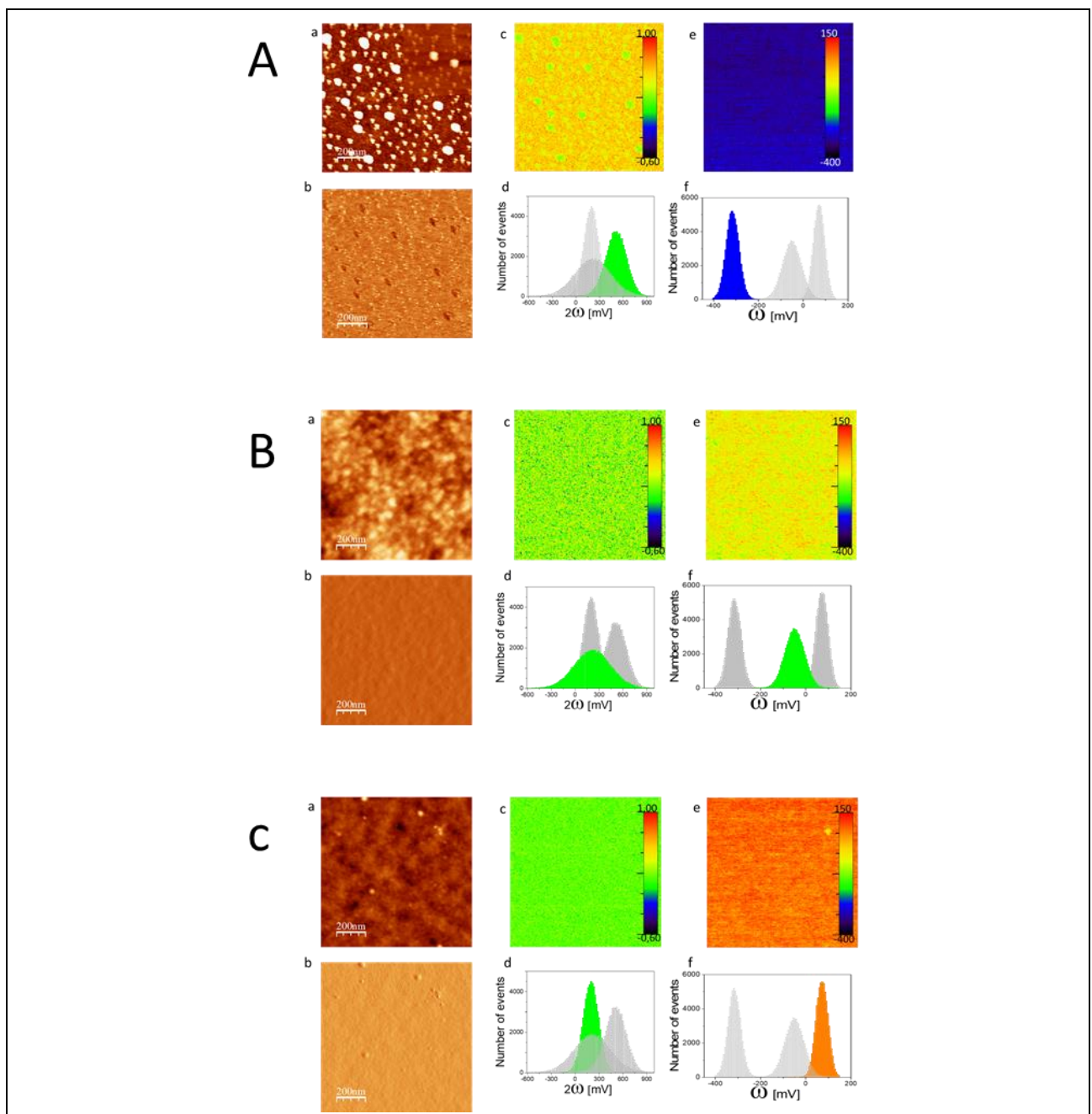

*Figure 2*

*Topography (a), error signal (b), capacity data (c) and contact potential (e) acquired on the flat part of three platinum coated AFM cantilevers: uncleaned-old (top, 2A), uncleaned new (middle, 2B) and cleaned (bottom, 2C). Lateral image size is 1x1 $\mu m^2$. Gray scale for the topography images is 5 nm (right top inset in figure 2Aa is 20nm), -0.6 - 0.1V for the contact potential image and ±0.8 V (arbitrary units, output of Lock-In Amplifier) for the capacity image. The histograms (d) and (f) show the variation of data in the corresponding images above; capacity (c) and contact potential, respectively. We note that, in order to avoid contamination from the sample to the tip the chronological order in which these images were acquired is: first the sample uncleaned–new, then (RCA) cleaned, and finally uncleaned–old.*

**Multidimensional AFM Spectroscopy of cleaned an uncleaned cantilevers**

To better characterize the tip sample system, we have acquired multidimensional microscopy data – so-called *interaction images* – at specific locations of the sample. These *interaction images* are then processed to separate Van der Waals and electrostatic interaction. For each interaction image three curves are obtained: a (true) Van der Waals interaction $\delta v_{VdW}(d)/v_0$ curve, the tip-sample capacitance $C''(d)$ curve (not shown) and a contact potential versus distance curve $U_{CP}(d)$. Essentially, from the capacity curve a radius $R$ is determined by fitting each curve to the second term in relation 3. From the Van der Waals versus distance curve (see graphs in Figure 3), the Hamaker constant is determined from a fit to the first term in relation 3 corresponding to the Van der Waals interaction (more precisely: first the product $A \times R$ is obtained from this fit, then with the tip radius $R$ obtained from the capacity curve the Hamaker constant $A$ is obtained).

Figure 3 shows the results of experiments where (topography) AFM imaging is combined with the acquisition of *interaction images* at well-defined spots. AFM data –topography images and *interaction images*– are acquired on samples of contamination type "*uncleaned–old*", "*uncleaned–new*", and "*cleaned*" (see discussion above). On each spot, several *interaction images* are obtained, the corresponding results are shown in the lower row of figure 3. For each single spectroscopy experiment n=1 - 15 two data sets are acquired, processed and shown: one corresponding to an approach cycle and another one to a retraction cycle of a spectroscopy experiment. Data sets n=1-5 are obtained on an *uncleaned old* cantilever, data sets 6-10 on an *uncleaned-new* and data sets n=11-14 on a (RCA) *cleaned* cantilever. All data is acquired in the true non-contact regime, that is, the tip-sample system does neither enter mechanical contact, nor are liquid necks formed. In fact, the oscillation amplitude during all experiments was constant, which implies no snap to contact nor sudden decrease of oscillation amplitude due to formation of liquid necks. Correspondingly the data is very reproducible and we observe a clear difference between data acquired on the sample with different degree of contamination. Clearly, the error determined from each experiment from the fitting procedure to the Van der Waals frequency curves is smaller than the difference of the Hamaker constant measured on the different cantilevers. We are therefore able to detect the variation of the Hamaker constant, as well as of the contact potential due to the chemical differences in induced by the contamination in the tip-sample system.

From the data shown in figure 3 we conclude that the effective Hamaker constant of the system decreases when the tip–sample system is more contaminated. This, in our opinion, is easily understood in terms of the higher Hamaker constant of platinum as compared to that of organic contaminants, which most likely constitute most of the contamination adsorbed on the tip and on the sample. In this context we note that our methodology gives the correct value of the Hamaker constant of the Platinum–Platinum system (about $A_{Pt}=240\times10^{-21}$ J) when the system has been (RCA) *cleaned* and immediately imaged. A lower, but still high (metallic–like) Hamaker constant (about $A_{Pt}=180\times10^{-21}$ J) is obtained for the cantilever of type *uncleaned-new*, and a (very) low Hamaker constant (about $A_{Pt}=80\times10^{-21}$ J) is obtained for a highly contaminated (*uncleaned-old*) tip–sample system. Interestingly, in this case essentially the same Hamaker constant is obtained on the flat part of the sample, as compared to the Hamaker constant measured on one of the drop like islands, which quite clearly correspond to contamination. We therefore conclude that on these latter samples (*uncleaned-old*) the contamination films should be quite thick; thick enough so that the interaction due to the Platinum on tip and sample is significantly smaller than the Van der Waals interaction of the contamination adsorbed, which gives a (relatively) high interaction force in spite of the lower Hamaker Constant because the contamination is closer as compared to the Platinum (see inset in figure 3).

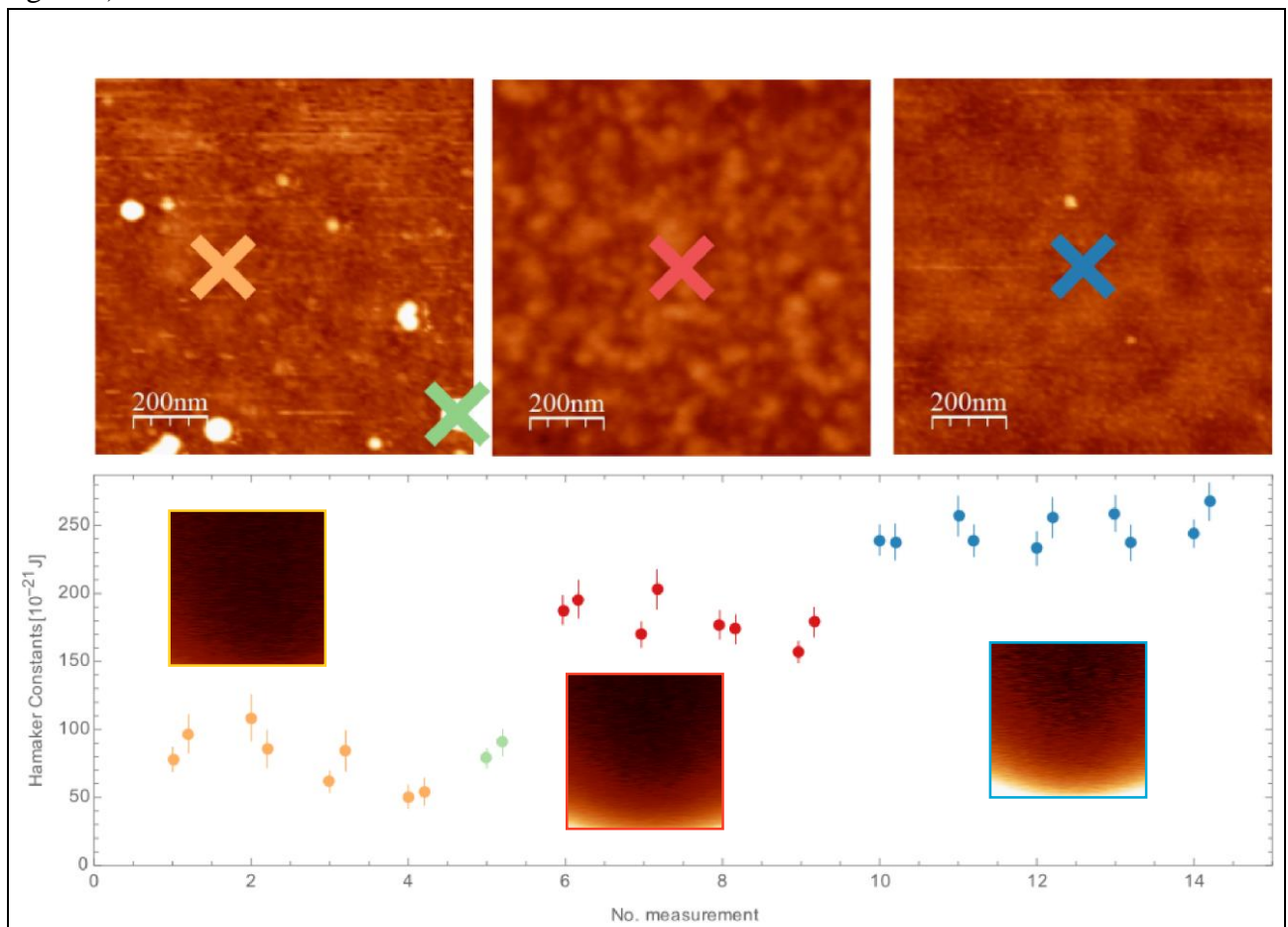

*Figure 3*

*Spectroscopy data on specific locations of three different samples with different degree of contamination (see main text): uncleaned–old (column A, left), uncleaned–new (column B, middle), cleaned (column C, right). Top row: topography images. Bottom row: Hamaker constant for the different interaction images acquired at the positions shown in the top row. The inset in these graphs show one of the interaction images (raw data, total gray scale 500 Hz) from which these Hamaker points are calculated.*

**Conclusions**

The present work describes our effort to shed light on the issue of contamination within an AFM tip–sample system. It is generally accepted that contamination must be present when working in ambient conditions; otherwise the extreme effort to work in UHV conditions would not make any sense. However, the precise way in which this contamination effects AFM data acquisition and data interpretation is – at the moment – not truly understood. In fact, in our opinion, too often this issue is just ignored in the AFM community. Since many more AFM related measurements are performed in ambient conditions as compared to UHV, and since AFM in ambient conditions has become a fundamental and widely used tool for nanoscale research in material science, physics, chemistry but also engineering sciences, we believe that this issue is not merely a "technical issue" related only to AFM operation, but a topic of broader relevance.

From our work we conclude that not only contamination of the tip, but also that of the sample is quite difficult to visualize and detect, but is nevertheless quite important, and significantly determines AFM results and interpretation. In the present work we have shown how the presence of contamination in the tip–sample system can be detected and quantified using only AFM techniques. The methodology presented allows for simple in–situ characterization and checking of the tip–sample system, without needing to remove the tip and/or the sample from the AFM set-up. To infer the contamination of the tip we have adopted the assumption that, if the whole cantilever is composed of the same material (as is the case for evaporated cantilevers) then the contamination should cover the tip apex, the tip and the flat part of the cantilever in a similar way. This assumption should be quite correct if the contamination layer is thinner than the radius of curvature of the tip, typically 20 nm, which –hopefully– should be the case in most AFM experiments. If this assumption is correct, taking two equivalent cantilevers and using one as probe and the other one as sample will allow to observe and measure the properties of the tip, since in a sense, the tip is looking at the (statistically) equivalent sample system. Within the Derjaguin Approximation (see relation 1) the chemistry of the tip sample system is described by the surface energy *w(d)* and we interpret that for a Pt-Pt tip-sample system this surface energy term is due to twice the same system: in our case Pt with a (dielectric) contamination layer. To validate our working assumption, we have imaged the tip and the flat cantilever regions of a cantilever taken from a typical enclosing box. As discussed previously in the literature, we find a high degree of contamination on the cantilevers and –more importantly in the present context– we observe that this contamination layer covers in a similar way the tip and the flat part of the cantilever (figure 1). In the second experiment we have used imaging of topography as well as of electrostatic properties to characterize cantilevers used as a sample. We have imaged three tip-sample systems with different degree of contamination, and find different electrostatic response of the capacity signal and of the contact potential (figure 2). Unfortunately, the data obtained from these images is not completely conclusive: the contact potential difference may indicate that the tip and the sample is either clean, or that both are contaminated in a similar way. Finally, we have shown that using multidimensional spectroscopy techniques (*interaction imaging*) and advanced data processing we are capable to a locally determine the effective Hamaker constant of the tip–sample system, and we find that this Hamaker constant is very sensitive to the contamination state of the tip–sample system (figure 3).

We propose the method presented in this work as a simple, very precise and sensitive, and -very important– as an in-situ characterization of the tip–sample system. In the future we think that this method can be easily extended to improve nanoscale material characterization in the following way: once the tip is known to be clean, the properties of an unknown sample can be characterized as discussed in the present work precisely because the tip side is clean and well controlled and not anymore an unknown parameter of the system. Only then the measured properties of the tip–sample system can be uniquely attributed to the sample. In the future we propose to further develop the technique presented here to make AFM a truly quantitative nanoscale Material Science characterization tool.

**Acknowledgements**

Research has been financed by the Ministerio de Ciencia e Innovación (MICINN, Spain) and the Ministerio de Economía y Competitividad through the projects "ForceForFuture" (CONSOLIDER program, CSD2010-00024), "Celulas Solares Nanoestructuradas fabricadas a partir de disoluciones: hacia una mejora de la eficiencia, estabilidad y escala del dispositivo" (ENE2013-48816-C5-1-R) as well as "Autoestructuración y Caracterización a la Nanoescala"(ENE2016-79282-C5-4-R).

**Notes and references**